\documentclass[a4paper]{llncs} 

\usepackage[dvips]{graphicx}
\graphicspath{{../eps/}{../IMG/}{./img/}}
\DeclareGraphicsExtensions{.eps, .pdf}

\usepackage{xspace}
\usepackage{xcolor}
\usepackage{amssymb}

\usepackage{array}
\newcolumntype{L}[1]{>{\raggedright\arraybackslash}p{#1}}
\usepackage{multirow} 

\usepackage{url}

\usepackage[utf8]{inputenc}

\usepackage[
backend=biber
,isbn=false
,url=false
,doi=false
,eprint=false
,maxbibnames=2
,firstinits=true
,alldates=year
,style=numeric
,sorting=none
]{biblatex}
\bibliography{bibSLR}

\urldef{\mailsa}\path|{fabian.kneer,|
	\urldef{\mailsb}\path|erik.kamsties}@fh-dortmund.de|
\urldef{\mailsc}\path|schmid@sse.uni-hildesheim.de|

\usepackage{color, colortbl}
\definecolor{Gray}{gray}{0.9}

\newcommand{\RQ}[1]{\textbf{[RQ{#1}]}}

\newcommand{\OR}{\texttt{OR}\xspace}
\newcommand{\AND}{\texttt{AND}\xspace}

\newcommand\parhead[1]{\vspace{2mm}\noindent\textbf{{#1}}\ \ }

\usepackage{times}
\usepackage{rotating}

\usepackage{enumitem}
\begin{document}
%
\title{Environment Modeling for Adaptive Systems: \\A Systematic Literature Review}

\author{Fabian Kneer\inst{1} \and Erik Kamsties\inst{1} \and Klaus Schmid\inst{2} }

\institute{Dortmund University of Applied Sciences and Arts,\\
	Emil-Figge-Str. 42, 44227 Dortmund, Germany\\
	\url{fabian.kneer@fh-dortmund.de}
	\mailsb\\
	\url{http://www.fh-dortmund.de/}\\
	\and
	University of Hildesheim,\\ 
	Universitaetsplatz 1, 31141 Hildesheim, Germany\\
	\mailsc\\
	\url{http://www.uni-hildesheim.de}
}

\toctitle{}
\tocauthor{}

	\maketitle

\begin{abstract}	
\textbf{[Context \& Motivation]} Adaptive systems are an important research area. The dominant reason for adaptivity in systems are changes in the environment. Thus, it is an important question how to model the environment and how to determine the necessary information on this environment in the requirements engineering phase.
\textbf{[Question/ Problem]} There is so far relatively little explicit study of the notion of environment models in software engineering research.
\textbf{[Principal ideas/ Results]} In this paper, we present a systematic literature review with the goal to determine the state of the art in environment modeling for adaptive systems, in particular from a requirements perspective.
We discuss the goals of the approaches,  the modeling concepts, as well as the methodology aspects of environment modeling in our survey. 
\textbf{[Contribution]} As major result of our survey, we provide a meta-model of existing environment modeling concepts. As a negative finding \textendash{} and a research opportunity \textendash{} we find that so far methodological aspects of environment modeling have received very little attention.

\end{abstract}

\begingroup
\let\clearpage\relax
\section{Introduction} 
\label{sec:intro}

The notion of adaptive systems, i.e., systems that react to observations and adapt their behavior accordingly has gained significant interest in the research community, due to its wide applicability~\cite{DeLemosGieseMueller+13}. The adaptation that happens in the context of adaptive systems is usually due to changes in the environment.\footnote{Though adaptations based on internal observations, e.g., in the context of self-monitoring, exist as well.} A requirements engineer needs information and knowledge about the environment to identify and develop possible adaptions of a system. The environment information need to be available for the engineers but also for the adaptive system itself in a formalized notation. Despite this need for formalized modeling of environments, there is so far relatively little explicit study on the notion of environment models in software engineering research.  

In this paper, we aim to change this situation. We performed a systematic literature review of environment modeling in adaptive systems.

Within this area we are interested both in ways to represent environment models as well as in methods to identify these environment models. This perspective is summarized in the following research questions:

\begin{description}[noitemsep,topsep=0pt,parsep=0pt,partopsep=0pt]
\item[\RQ{1}] What types of  environment models and respective concepts are used to represent environment models for adaptive systems in the requirements engineering phase?
\item[\RQ{2}] What methods are used to identify the relevant information for environment models in the requirements engineering phase?
\end{description}

As basically all research that we could identify describes a modeling approach, while only very few papers address the methodology aspect as well, the main part of our contribution will be a systematic overview of the modeling approaches typically used for environment modeling.

Modeling the environment is also discussed in the general RE literature, i.e., outside the realms of adaptive systems, e.g., \cite{Pohl2010}. There typically  a variety of entities such as users, sensors, actuators, documents, laws) are described. As adaptive systems \textit{monitor} the environment, we expect to see:
 
\begin{itemize}[noitemsep,topsep=0pt,parsep=0pt,partopsep=0pt]
	\item a precise, focused understanding of the notion of environment, i.e., what are the environment entities relevant to adaptive systems?
	\item dedicated environment models, e.g. provided by a meta-model of the constituents and their relationships,
	\item specialized constructs in an environment model relevant to adaptive systems,
	\item a methodological treatment of the former points.
\end{itemize}

The results of our study may help to cross-fertilize the different views on the environment in different approaches and help to come up with a common understanding of environment modeling for adaptive systems. 
  
The remainder of this paper is structured as follows: in Section~\ref{sec:Method}, we describe our approach for executing the systematic literature survey. We discuss both how we identified the relevant sources as well as how we extracted and systematized the information from the papers. Section~\ref{sec:Findings} summarizes our findings and discusses some particularly relevant  results. Section~\ref{sec:relatedWork} presents the related work and discusses the differences to our study. In Section~\ref{sec:ThreatsToValidity} we present the threats to validity and how we handled these threats.
Finally, Section~\ref{sec:Conclusion} provides the conclusion.

\section{Method}
\label{sec:Method}
This section describes the method used in our systematic literature review. The structure and process is based on the guidelines provided by Kitchenham et al.~\cite{Kitchenham07guidelinesfor}.  

It is structured as follows: we describe  the search strategy in Section~\ref{subsec:SearchStrategy},  including the databases used and the derived search strings. The selection criteria for screening the papers are given in Section~\ref{subsec:SelectionCriteria}, while Section~\ref{subsec:DataExtraction} describes the data extraction process and the documentation of the results.

\subsection{Search Strategy}
\label{subsec:SearchStrategy}

Our systematic literature review uses a systematic derivation of search strings and several electronic databases for identifying the relevant literature. 

\parhead{Electronic Databases.} The following electronic databases were used as resources in order to identify and collect relevant material: IEEE Xplore Digital Library\footnote{\url{http://ieeexplore.ieee.org}}, Springer Link\footnote{\url{http://link.springer.com}}, Google scholar\footnote{\url{https://scholar.google.de}}.

All three databases support full text and meta-data search. IEEE Xplore and SpringerLink focus on publications of the respective publishers which resulted in a insufficient amount of publications. Thus, we used Google Scholar as a third resource to ensure a sufficiently broad coverage of all relevant sources.

\parhead{Search Strings.}  In order to derive the search strings, we used a basic pattern with three facets. The first is the object of our research: the environment model, the second is a qualification: adaptive and the third is our restriction, the focus on requirements engineering. By identifying alternative wordings for each of these three facets, we determined the final search string. Different alternative wordings were tested to find the best balance between accuracy of the resulting set of publications and feasibility of the study.  

\begin{itemize}[noitemsep,topsep=0pt,parsep=0pt,partopsep=0pt]
\item 1-facet: environment model and environmental model. 
\item 2-facet: adaptive, adaptation, self-adaptive, self-configuring, self-reconfiguration, self-optimization, and self-awareness. 
\item 3-facet: requirements engineering.
\end{itemize}

The following search string is produced by combining these facets along with their alternative terms:

\noindent ("\textit{environmental model}" \OR "\textit{environment model}") \AND
(\textit{adaptive} \OR \textit{adaptation} \OR \textit{self-adaptive} \OR self-configuring  \OR \textit{self-reconfiguration} \OR \textit{self-optimization} \OR \textit{self-awareness}) \AND 
("\textit{requirements engineering}")

The three databases and the search string are used during the selection process. This process is described next.

\subsection{Selection Criteria}
\label{subsec:SelectionCriteria}

A number of criteria were used for making the selection reproducible, which are given below.

\parhead{Inclusion Criteria.} Only peer-reviewed workshop, conference, and journal papers were accepted, which (1) are from the areas of requirements engineering and self-adaptive systems, (2) include a textual or graphical model to define or describe the environment of a self-adaptive system, or (3) describe the extension of a model to define or describe the environment of a self-adaptive system. 

\parhead{Exclusion Criteria.}
Papers are excluded if they meet one of the following criteria: papers and reports have been excluded where only the abstract, but  not the full text is  available. Publications have been excluded if they are not written in English. Letters and editorials have been excluded as well as learning materials and theses. Books have been excluded and papers under five pages in length.

The exclusion of papers takes place after the articles have been ordered and the meta-information that is needed for the selection process is stored.

\parhead{Selection Process.}
To identify the relevant literature for our review, we used an automated search and then applied the criteria to include or exclude a publication. This process consists of three steps:

\noindent Step 1: Collect and store papers from the chosen databases. The previously defined search strings are used to automatically search for relevant papers. 

\noindent Step 2: The titles and abstracts of identified papers are screened for relevance to the topic. We read the papers and search for violations of the previously defined inclusion or exclusion criteria.

\noindent Step 3: The full text of papers was analyzed for all studies appearing to meet the inclusion criteria and a final selection was made.

The numbers of included and excluded publications are recorded during the different stages of the search process. We started with 455 publications (Springer 98, IEEE 32, Google Scholar(GS) 325). During Step 2 the total number are reduced to 226 (Springer 66, IEEE 20, GS 140). After Step 3, the selection process resulted in 76 paper (Springer 24, IEEE 10, GS 42). 18 of the 76 papers were identified as duplicates in two of the databases. We ended up with 58 accepted publications for the literature review.

\subsection{Data Extraction}
\label{subsec:DataExtraction}

During the systematic literature review, Citavi\footnote{\url{https://www.citavi.com/en}} was used for managing the 
collected references.
For each included publication the bibliographic information (title, DOI, etc.) was recorded, along with the abstract and a short note on why the paper was included.
The full text of the included papers was stored in a git repository that was used for sharing the results during the search process.
The publication in the reference manager and the file in the repository share the same ID (Bib\_Key).

In addition to the meta-information, the following information was collected from the selected publications: time stamp of the data extraction, publication type (journal, conference, workshop, etc.), aim and/or objectives of the publication, name of the used environment model, characteristics of the model (elements, links, textual/ graphical, etc.),  techniques to elicit the environment (if any), validation method, findings, and conclusion. The three authors of this paper read the selected publications and extracted all important information for the systematic literature survey.

\section{Findings}
\label{sec:Findings}

In this Section, we present the findings from our systematic literature review. First, the results related to the two research questions are represented in Section \ref{subsec:RQ1} (\RQ{1}) and \ref{subsec:RQ2} (\RQ{2}), followed by an Interpretation in Section \ref{subsec:Interpretation}. The results are interpreted and major findings are highlighted.

\subsection{\RQ{1} What types of environment models and respective concepts are used to represent environment models for adaptive systems in the requirements engineering phase?}
\label{subsec:RQ1}

\vspace*{-2.5em}
\begin{table}
	\centering
	
	\caption{Modeling approaches used in literature}
	\label{tab:Models}
	\tabcolsep=2pt
	\rotatebox{90}{
		\resizebox*{0.8\textwidth}{!}{
			\small
			\begin{tabular}{|l|L{6cm}|L{5cm}|}
				\hline 
				& Modeling Approach &  Paper\\
				\hline
				\multirow{7}{*}{\rotatebox{90}{Agent-based}} & Agent	&  \textbf{\cite{Choren.2005c}~\textsuperscript{*}}, ~\textbf{\cite{Choren.2005b}~\textsuperscript{*}}, ~\cite{Cossentino.2010}, ~\textbf{\cite{Liu.2014}~\textsuperscript{+}}, ~\textbf{\cite{LopezLorca.2016}~\textsuperscript{+}}, ~\textbf{\cite{LopezLorca.2016b}~\textsuperscript{+}}, ~\textbf{\cite{Mellouli.2004}~\textsuperscript{+}}, ~\cite{Mili.}, ~\cite{Mili.2008}, and ~\textbf{\cite{Athanasiadis.2009}~\textsuperscript{-}}\\ 
				
				& ROADMAP&  \cite{Juan.2004}\\
				
				& VigilAgent &  \cite{Gascuena.2014}\\
				
				& UML class diagram notation for Agents&  \cite{HendersonSellers.2004} and~\cite{Giese.2007}\\
				\cline{2-3}
				
				& Bio-inspired Task Model &  \cite{LopezJaquero.2016}\\
				\cline{2-3}
				
				& Capability Model&  \cite{Stirna.2012}\\
				
				& Enterprise Modeling (EM) and Capability Modeling&  \cite{Berzisa2015}\\
				\hline
				\multirow{5}{*}{\rotatebox{90}{Goal-based}} & Goal & \cite{Ceret.2013}, ~\cite{Kim.2009}, ~\textbf{\cite{Liu.2014}~\textsuperscript{+}}, ~\textbf{\cite{LopezLorca.2016}~\textsuperscript{+}}, ~\textbf{\cite{LopezLorca.2016b}~\textsuperscript{+}}, ~\textbf{\cite{Mellouli.2004}~\textsuperscript{+}}, ~\textbf{\cite{Qureshi.2012}~\textsuperscript{\#}}, ~\cite{Jian.2010}, ~\cite{Jureta.2008} and~\textbf{\cite{Baresi.2013}~\textsuperscript{\$}}\\
				& i* &  \cite{Carvallo.2009},~\cite{Pimentel.2012}, and~\cite{Sawyer.2007}\\
				& KAOS &  \cite{Fredericks.2016},~\cite{Giese.2014b}, and~\cite{Nakagawa.2011}\\
				& Tropos & \cite{Morandini.2015} \\
				& NFR Pattern in i* &  \cite{Cunha.2014}\\
				
				\hline
				\multirow{4}{*}{\rotatebox{90}{Rule-based}}& Domain Assumption&  \textbf{\cite{Baresi.2013}~\textsuperscript{\$}}\\
				& Rule-based & \cite{Wang.2015} \\
				& Planning with Action Prioritization (PAP) &  \cite{Ghosh.2015}\\ 
				& CONSENS & \cite{Holtmann.2016}\\
				
				\hline
				\multirow{10}{*}{\rotatebox{90}{State machine}} & State machine & \cite{Lu.2008} \\
				& Stochastic Hybrid Automata (SHA)&  \cite{Chen.2015}\\
				& Stochastic Timed Automata (STA)&  \cite{Weyns.2016}\\
				
				& Labeled Transition Systems (LTS)&  \cite{Dippolito.2011}\\
				& Labeled Transition Kripke Structure (LTKS)&  \cite{DIppolito.2014}\\
				& Linear Temporal Logics (LTL) &  \cite{DIppolito.2010} and~\cite{DIppolito.2013}\\
				
				& Bigraphs&  \cite{Yu.2014} and~\cite{Yu.2016}\\
				& Petri net&  \cite{Han.2012}\\
				
				& Internal structure and behavior model &  \textbf{\cite{Athanasiadis.2009}~\textsuperscript{-}}\\
				& Time Usage Models (TUM)&  \cite{Siegl.2015}\\
				\cline{2-3}
				& RTES Environment Modeling &  \cite{Iqbal.2012b} and~\cite{Iqbal.2015b} \\
				& Behavior & \cite{Uhrmann.2012}\\
				\hline
				\multirow{5}{*}{\rotatebox{90}{Other}}
				& Ontology&  \textbf{\cite{Choren.2005c}~\textsuperscript{*}},~\textbf{\cite{Choren.2005b}~\textsuperscript{*}},~\cite{Qureshi.2011}, \textbf{\cite{Qureshi.2012}~\textsuperscript{\#}}, and~\cite{Sharifloo.2015}\\
				\cline{2-3}
				& Soft Systems Methodology (SSM) \& Viable System Model (VSM) &  \cite{Bustard.2005}\\
				& Environment Context Model & \cite{Ghannem.2015} \\
				& Digital Virtual Worlds&  \cite{Pechoucek.2012}\\
				\cline{2-3}
				& Not specified & \cite{Axenath.2006},~\cite{Kromker.2015},~\cite{Abuseta.2015},~\cite{Dignum.2002}, and~\cite{Holtmann.2015}  \\
				\hline
			\end{tabular} 
		}
	}
\end{table}
\vspace*{-1em}

Related to RQ1 we identified the modeling approaches shown in Table~\ref{tab:Models}. We recognized four major types of models: \textit{agent-based} 17 (29\%), \textit{goal-based} 18 (31\%), \textit{rule-based} 4 (7\%) and \textit{state machines} 15 (25\%). 
In cases, in which an approach makes use of a particular combination of modeling paradigms, we indicate this by a symbol in superscript at the respective entries in Table \ref{tab:Models}. For example, all approaches that combine agent and goal modeling are marked with the superscript \textsuperscript{+}, e.g. Lopez-Lorca et al.~\cite{LopezLorca.2016}.

Agent- and goal-based approaches mostly represent environment entities using resource elements that can be monitored or modified. Additionally, some authors define the environment using actors or agents and describe the abilities or resources in the same way as other actors in the system.

Rule-based approaches make use of rules that consist of events, conditions, and actions. These rules represent the environment by an event that appears in the environment, which results in an action of the system or in an action of a controllable element in the environment.

State machine approaches describe the behavior of the environment by using states and transitions to represent different environment situations and how the environment changes between these situations.

14  approaches (24\%) are included in the \textit{other} category. These works created their own modeling approach, like Ghannem et al.~\cite{Ghannem.2015}, who focus in their contribution on the development of an environment model for a specific domain (parking lot management).  Ontology approaches model the environment using an ontology. Qureshi et al.~\cite{Qureshi.2012} for example, propose an ontology based on a goal-based modeling approach, see Table \ref{tab:Models} superscripts~\textsuperscript{\#}.

The publications in the \textit{not specified} area are vision papers, "Towards..." papers, or papers with an unclear description of the environment like a box in a diagram with a brief explanatory text. 

The different concepts of the approaches and their relationships will be discussed later.

\noindent\fbox{\parbox[t]{\dimexpr1\linewidth-2\fboxsep-2\fboxrule\relax}{
	\textbf{Finding 1:} \textit{The major modeling approaches are goal-based (31\%), agent-based (29\%), state machines (25\%), and rule-based (7\%).}	
}} 

Below, we discuss  the different modeling approaches using the criteria that we collected during the data extraction (see Section \ref{subsec:DataExtraction}).Base on the data we propose a meta-model that summarizes the concepts of environment models that we identified in the publications.

\parhead{Representation of the environment model.}
The next criterion is the representation of the environment model. Figure~\ref{fig:graphictextual} shows that 48\% are pure \textit{graphical} models, which consist of 2\% \textit{3D} modeling and 46\% ordinary \textit{graphical} models. 24\% make use of a combination of graphical models and textual descriptions. It is the sum of \textit{Both} 14\%, \textit{Graphical Tabular} 8\%, and \textit{Graphical (Annotation) Tabular} 2\%. 10\% are pure \textit{textual} descriptions, 6\% are \textit{tabular} descriptions to structure the textual information, and 12\% are not specified. In total, 72\% of the approaches use a graphical representation compared to 16\% that use a textual representation.

\begin{figure}[!t]
	\centering
	\includegraphics[width=0.85\linewidth]{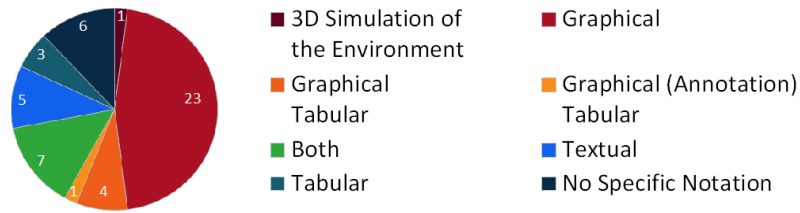}
	\caption{Representations of Environment Models}
	\label{fig:graphictextual}
\end{figure}

One motivation for combining textual and graphical representations is to use the best of both worlds and to avoid clutter and information overload. For example, Jian et al.~\cite{Jian.2010} use i* models to represent the adaptive system and use tables with additional information for an activity (task), like for instance an environment element, trigger, or condition. This information is used to combine an activity with the environment and to react on changes. Another combination of graphical and textual representations can be seen in the approach of Pimental et al.~\cite{Pimentel.2012}. They defined annotations for a goal model, which identify relations to the context. These relations are explained in natural language  in a table.

Textual approaches describe the environment with textual elements that follow a pattern. These descriptions are mostly structured in a tabular form. For example, Kim et al.~\cite{Kim.2009} use tables to structure goals and scenarios, which in turn are used to define states and actions in another table.

The last representation is a virtual 3D real world simulation of the environment by P{\v{e}}chou{\v{c}}ek~\cite{Pechoucek.2012}. In their "Towards..." paper, the authors describe the need for a simulated environment to develop and validate multi-agent systems in an early phase of the development.

\noindent\fbox{\parbox[t]{\dimexpr1\linewidth-2\fboxsep-2\fboxrule\relax}{ 
		\textbf{Finding 2:} \textit{72\% of the approaches use a graphical visualization and 16\% use textual descriptions.}	
}}

\parhead{Meta-Model of existing environment modeling approaches.}
One of our major results is a meta-model of concepts for environment modeling, which is shown in Figure~\ref{fig:ele_all}. All concepts were extracted from the accepted publications. Out of this list we combined concepts with the same meaning. For example, terms like \textit{environment}, \textit{environment entity}, \textit{environment context} are combined to \textit{environment entity}. Another example is \textit{monitored entity} which refers to terms like \textit{sensor}, \textit{measurable property}, \textit{input}, and \textit{context indicator}. 
\begin{figure}
	\centering
	\includegraphics[width=0.8\linewidth]{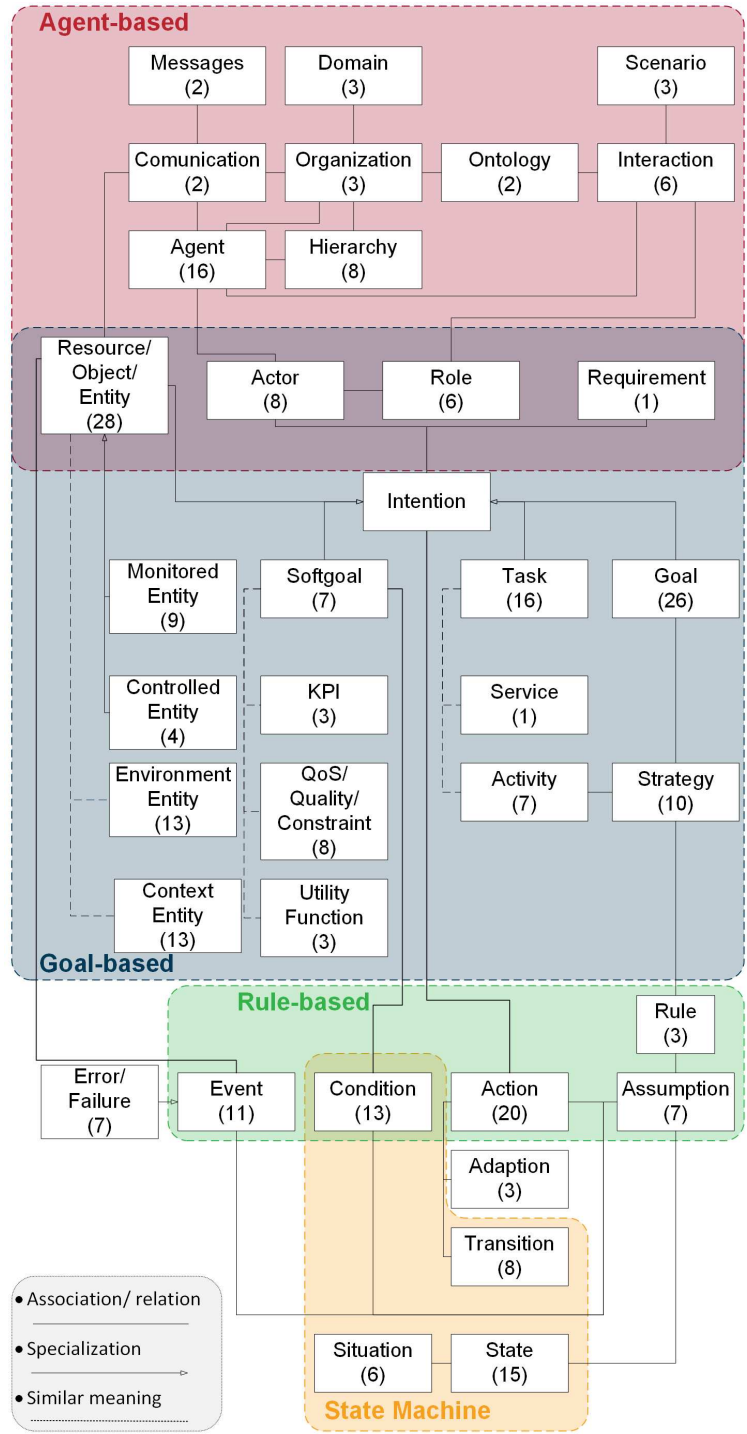}
	\caption{Meta-model of existing environment modeling approaches}
	\label{fig:ele_all}
\end{figure}

The \textit{monitored-} and \textit{controlled} entities are specializations for an environment entity, which can be monitored, like a sensor or controlled, like an actuator. 

We identified four clusters of concepts, namely agent-based, goal-based, rule-based, and state machine concepts. The clusters are connected, i.e., they share concepts. Also many approaches mix these techniques, for example, goal- and agent-based by Lopez-Lorca et al.~\cite{LopezLorca.2016} and~\cite{LopezLorca.2016b}. 

Concepts of the state machine and rule-based cluster are often used together with concepts of the agent- or goal-based cluster. As this survey is focused on adaptive systems, these systems need to change after changes in the environment happen. That is, the event that triggers a rule or a state transition results from a change in a resource, such as a monitored entity (e.g., sensor). The conditions are based on quality constraints like QoS or softgoals. These concepts are used in a similar way but can not be combined in a single concept (connected in the meta-model by dashed line). 

The resulting action can be a state transition or an adaptation of any intention, like a goal, a task, or service or a change in the environment (controlled entity).

The goal of the meta-model is to highlight the connections among the concepts of the different models. Therefore, we excluded some concepts to make the model more general and to avoid overload. These concepts are specific to the domain or system, for instance, aisle, room, furniture, door, location, measurements. Data, parameter, class, attribute, capacity, property are examples for general concepts.

\vspace*{1ex} 
\noindent
\fbox{\parbox[t]{\dimexpr1\linewidth-2\fboxsep-2\fboxrule\relax}{
		\textbf{Finding 3:} \textit{A meta-model of existing modeling concepts and the connection among the concepts (Figure~\ref{fig:ele_all}). In this model we combine different terms with the same meaning to generate a general overview of the concepts of environment modeling. Also, we show direct references and  concepts shared among the major approaches.}	
}}\vspace*{1ex}

\subsection{\RQ{2} What methods are used to identify the relevant information for environment models in the requirements engineering phase?}
\label{subsec:RQ2}
Figure~\ref{fig:methodology} shows the results related to \RQ{2}. The selected papers contribute to design (39), modeling (39), validation (9), and elicitation (9). That is, the publications mainly focus on the development of self-adaptive systems. In particular, they focus on the design of architecture and on modeling of the environment. These results are expected as we searched for environment models. 
\begin{figure}[!t]
	\centering
	\includegraphics[width=0.75\linewidth]{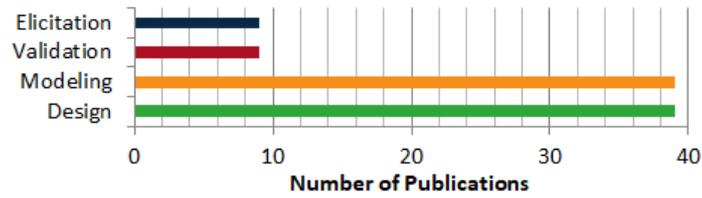}
	\caption{Methodological aspects of Environment Modeling for Requirements Engineering}
	\label{fig:methodology}
\end{figure}

Unexpected for us are the few results for validation and elicitation of the environment. Elicitation is addressed mostly with a few sentences that say the environment information is elicited using interviews or meetings, like Athanasiadis et al.~\cite{Athanasiadis.2009} and Gasecuena et al.~\cite{Gascuena.2014}, or implicitly through the development of a goal model and scenario descriptions, like Carvallo et al.~\cite{Carvallo.2009} and Choren et al.~\cite{Choren.2005b},~\cite{Choren.2005c}.

Only one paper focuses on elicitation and compares different elicitation methods like Brainstorming, 6-3-5 method, six thinking hats, field observation, apprenticing, and interviews. We expected a bigger focus on field observation to elicit the environment as information about the environment is a priori often not available.
Unexpected were also the results for validation. One author addresses the validation of systems with an environment model and the validation of the environment model is discussed in one paper only. Siegl et al.~\cite{Siegl.2015} discuss properties like completeness, traceability, correctness, and consistency of the environment model.

Most other authors used the model to simulate the environment and to validate and verify the behavior of an adaptive system. For example, P{\v{e}}chou{\v{c}}ek et al.~\cite{Pechoucek.2012} use a mixed-mode simulation. That is, the test system runs in a mix of real system parts and simulated virtual environment.

\vspace*{1ex} 
\noindent\fbox{\parbox[t]{\dimexpr1\linewidth-2\fboxsep-2\fboxrule\relax}{
		\textbf{Finding 4:} \textit{The focus is on the development and design of adaptive systems. Only one paper discusses elicitation in detail and one paper discusses important qualities of an environment model.}	
}}
\vspace*{1ex}

\subsection{Interpretation of \RQ{1} and \RQ{2}}
\label{subsec:Interpretation}
In this subsection we interpret the results with respect to our expectations (discussed in Section \ref{sec:intro}). We found examples of environment entities that are relevant to adaptive systems. For example, the heating system used by \cite{Chen.2015}, refers to additional environment information of a room, like wall thicknesses and sizes, to make better decisions. These properties are not monitored or controlled by the system, but they are required to adapt the system to a particular context. However, we did not find a generalized notion of environment in the context of adaptive systems.

We expected a dedicated environment model which captures all the environment information relevant to an adaptive system. Yet, we found mostly approaches, which annotate system models to represent the environment conditions (or  contain only parts of the environment). For example, in the goal- and agent-based approaches the environment can be represented as a single concept mostly called \textit{resource} or as an \textit{actor} that contains the modeling concepts, which are  used also to represent a system.

As we expected, some approaches added specific concepts to the environment model, such as a specialization of a \textit{resource} into a \textit{controlled} or \textit{monitored} variable. These specific elements help to separate system and environment resources and also to add more details about the characteristics and capabilities of the environment. 
\section{Related Work}
\label{sec:relatedWork}

We identified two literature studies which focus on RE for self-adaptive systems. The first was a systematic literature review  by Yang et al. \cite{Zhuoqun2014}. Their aim was to identify modeling methods, RE activities, requirements quality attributes, application domains, and research topics in the area of self-adaptive systems. The study consists of 101 relevant papers. 

The second was a mapping study performed by Sucipto et al.~\cite{Sucipto_2015}. Similar to the first study, their goal was to summarize the research topics, categorize the used modeling methods and requirements activities for the development of self-adaptive systems. They accidentally identified 101 relevant paper as well. The authors report that only 3 out of 101 papers cover elicitation activities which is backed by our results.

Compared to our SLR, both studies are much broader in their scope. With respect to our research question RQ1, these studies just name modeling techniques used for adaptive systems. We add to these general findings by providing deeper insights into environment models and their concepts in particular.

\section{Threats to Validity}
\label{sec:ThreatsToValidity}
A threat to validity of our analysis is a personal bias, as the analysis was mostly performed by the first author of this paper. Additional randomized analyses were done by the second and third author to identify any issues.  This did not uncover major issues. The search strategy and findings were discussed and agreed upon among all authors.

A second threat is that the results depend on the employed databases. In order to address this threat, we used three different databases with a different, while overlapping coverage. All searches with their results were archived to enable later cross checks. 

The selection of keywords imposes a third threat. First, we tried to find a balance between accuracy of the resulting set of publications and feasibility of the study. A problem in this regard was caused by the term \textit{context}, which is used frequently (but not always) synonymously to \textit{environment}. The term \textit{context} in itself bears a significant vagueness (as it is used also in the sense of \textit{discourse} or \textit{setting}). We tried to include \textit{context} in the search string and also used different alternatives for context, which resulted in sets of 5000 to 20000 publications. As this was significantly outside the acceptable range and caused a huge number of false positives, we ultimately decided to remove this from the search string. 
Second, we ensured the quality of the keywords by identifying five publications that we consider as relevant to the focus of the study. Then we applied the search strings and checked if these 5 publication were selected. In relation to the third threat, the chosen keywords and resulting set of publications contains theses five publications. 
\section{Conclusion} 
\label{sec:Conclusion}

Research on adaptive systems received increasing interest over recent years. An important topic is the environment model, as this plays a central role in every approach to adaptive systems that addresses changes in the environment. 

In our  systematic literature review, we found a significant amount of research on environment modeling with an increasing trend.  Overall, 455 publications were collected from different databases and eventually 58~publications were selected as relevant in a systematic and repeatable process described in the paper.
 
From these publications we identified four major modeling approaches: agent-based, goal-based, rule-based, and state machines. Additionally, notational concepts were identified,  analyzed, and combined into a joint meta-model. This meta-model shows the connections among the approaches by using joint modeling concepts, e.g., \textit{actor}. 
Another contribution is a categorization, which shows the interests and goals in the area of  environment modeling for adaptive systems.

We identified several research opportunities. First, we found only one publication that focuses on the elicitation of the environment model.  Also the validation of these models in terms of completeness, traceability, correctness, and consistency is an open topic. Beyond this, various research opportunities exist, which are related to the fact that very different (even conceptually non-overlapping) modeling approaches for environment modeling are used.   
\endgroup

\printbibliography[notkeyword=SLR]
\printbibliography[keyword=SLR, title={SLR References}]

\end{document}